# Inner-shell Annihilation of Positrons in Argon, Iron and Copper Atoms

**M. A. Abdel-Raouf(1), M. M. Abdel-Mageed(2) and S. Y. El-Bakry(2)**

**(1)Physics Department, Faculty of Science, UAEU, Al Ain 17557, United Arab Emirates**

**(2)Physics Department, Faculty of Science, Ain Shams University, Cairo, Egypt**

# Abstract

**The annihilation parameters of positrons with electrons in different shells of Argon, Iron and Copper atoms are calculated below the positronium (Ps) formation thresholds. Quite accurate ab initio calculations of the bound state wavefunctions of Argon, Iron and Copper orbitals are obtained from Cowan computer code. A least-squares variational method (LSVM) is used for determining the wavefunction of the positrons. The program is employed for calculating the s-wave partial cross sections of positrons scattered by Iron and Copper atoms. Our results of the effective charge are compared with available experimental and theoretical ones.**

---

assad@uaeu.ac.ae

## 1-Introduction

Positrons, since their prediction by Dirac [1], and their observation by Anderson [2] and Blackett and Occhialini [3], have been extensively used as probes in different branches of physics. Studies of positron interactions with gases, atoms and molecules with and without positronium formation have been carried out over the years. Much interest, however, has been devoted in the last decade to the interaction of slow positrons (below the positronium formation thresholds) with one, two and three dimensional macromolecular structures, (e.g. large molecules, chains, surfaces, crystals and bulks). In these cases, positron annihilation is employed as an effective nondestructive tool for the investigation of electronic structures as well as defects of materials. Particularly, positrons may also annihilate with inner-shell electrons creating holes which consequently induce the emission of highly efficient Auger electrons with extremely low background secondary electrons as the ultimate parallel tools (see e.g. Weiss et al [4] and Kim et al [5]) for the investigation of surface metals.

The aim of the present paper is to use a least-squares variational method (LSVM) for determining the wavefunctions of slow positrons interacting with inner-shell electrons of different atoms below the positronium (Ps) formation thresholds. In order to illustrate the strength of our algorithm, we study the annihilation of positrons with one of the noble gases, namely Argon. (Previous interesting works on positron collisions with noble gases were carried out by Montgomery and LaBahn [6] and McEachran et al [7]). Our main goal, however, will be the annihilation of positrons in two, from the industrial point of view, extremely important metals, namely iron and copper. Each wavefunction is used to calculate the effective charge (annihilation parameter) $Z_{eff}$, which stands for the effective number of electrons at the positron position at a given subshell of the target atom. In this case the calculation of annihilation rates and cross sections are directly related to the average density of electrons at the position of the positron. An elaborate version of Cowan computer code ([8], program RCN32) is used to calculate quite accurate ab initio orbital wavefunctions of the target atoms.

## 2-Theory

In non-relativistic time-independent quantum mechanics, Schrödinger's equation is written as

$$(H - E)\,\Psi = 0 \;, \qquad (1)$$

where H and E are the total Hamiltonian and energy, respectively, of a quantum mechanical system described by the wavefunction $\Psi$. The boundary conditions of $\Psi$ characterize various quantum mechanical systems, e.g. bound-state system, scattering process, etc. In the collision of positrons ($e^+$) with target atoms (A), the positrons annihilation is subjected to the emission of 2 or 3 $\gamma$ photons according to one of the following processes:

$$e^+ + A \rightarrow \begin{cases} A^+ + 2\gamma \ (\text{or } 3\gamma) \quad (\text{Direct annihilation}) \\ Ps + A^+ \rightarrow A^+ + 2\gamma \ (\text{or } 3\gamma) \\ [e^+, A] \rightarrow A^+ + 2\gamma \ (\text{or } 3\gamma) \end{cases} \qquad (2)$$

Ps and $A^+$ stand for the positronium and the residual ion. In the first process which is called direct annhilation, the incident positron annihilates ( below the Ps formation threshold ) with one of the atomic electrons of the neutral target atom A and the annihilation rate is calculated using the electron charge density ( $Z_{eff}$ ) at the positron position. In the second process the incident positron (above the Ps formation threshold) picks up an electron to form positronium and after that annihilates. The positron in the third process is captured to the atom to form [$e^+$,

A] bound system and the photon annihilation then occurs within the positron-many-electron complex system.

In the present work, we concentrate ourselves on the first (i.e. the direct annihilation) process. In positron-atom scattering, $\lambda$ , the rate of annihilation of an incoming positron and an atomic electron with the emission of two gamma rays, is given by the expression (Ferrell [9] and Fraser [10])

$$\lambda = \pi \, r_0^2 \, c \rho \, Z_{eff}(k) \quad , \tag{3}$$

where $r_0$ is the classical radius of the electron, c is the velocity of light. $\rho$ is the density of electrons per atom available to the positron for annihilation and $k$ is the positron wave number. $Z_{eff}(k)$ is defined in general as the effective number of electrons per atom available to the positron for annihilation. (In our case it stands for the effective number of electrons occurring at the positron position at a given subshell of the target atom). It depends on specific properties of the $e^+$ - Atom system under consideration and is equal to $Z$ , the number of atomic electrons, if the interaction potential between the positron and the atom is set to be zero. The annihilation parameter $Z_{eff}(k)$ can be calculated using the scattering wavefunction obtained via the least-squares variational method. Remembering that the annihilation parameter is related to the probability of an electron and a positron to be found in the same position, we can write

$$Z_{eff}(k) = \left\langle \Psi(x,r;k) \right| \sum_{i=1}^{N} \delta(r_i - x) \left| \Psi(x,r;k) \right\rangle, \tag{4}$$

where $\Psi\left(x,r;k\right)$ is the full scattering wavefunction, including all partial waves, for the system made up of the incident positron with wave vector **k** and the target atom. **x** and **r** stand for the position vector of the positron and the target (composed of N electrons), respectively. For s-wave scattering process, the variational treatment (Abdel-Raouf [11]) starts by defining a trial wavefunction $\Psi_t^n\ (x,\ r;k)$. It consists of two multiplicative wavefunctions

$$\Psi_t^n\ (x,\ r;k)\ =\ \Phi_T(r)\ \Psi_{Sc}^n(x;k) \tag{5}$$

where $\Phi_T(r)$ represents the target in its ground-state and $\Psi_{Sc}^n(x;k)$ is the positron scattering wavefunction which is composed of the angular part ( $Y_{0,0}=\sqrt{1/4\pi}$ ) multiplied by the radial part $\Psi_P^n(x;k)$. Thus, we have

$$\Psi_P^n(x;k) = a^n\hat{S}(x;k)\ +\ b^n\hat{C}(x;k) + \sum_{i=1}^{n} d_i\chi_i(x), \tag{6a}$$

n refers to the dimension of the square integrable part of the trial wavefunction representing all possible virtual states of quantum mechanical system composed of the positron and the target. $\hat{S}(x;k)$ and $\hat{C}(x;k)$ specify the regular and irregular parts of the wavefunction, respectively. Usually, the latter is accompanied with a cut-off function for avoiding the singularity at the origin. This cut-off function will tend to zero at the origin and to unity at infinity. $\Psi_P^n(x;k)$ has to satisfy the boundary conditions:

$$\Psi_P^n(0;k)\ =\ 0$$

$$\Psi_P^n(x;k)\ \xrightarrow{x\to\infty}\ a^n\,\hat{S}(x;k) + b^n\,\hat{C}(x;k) \tag{6b}$$

The function $\chi_i\,(x)$ appearing at eq. (6a) is a square integrable wavefunction. $a^n$ , $b^n$ *and* $d_i$ are variational parameters. In this case the reactance matrix K contains a single element which is identical with the tangent of the s-wave scattering phase shift ( $\eta_0$ ) and is calculated by

$$K_{11} = \tan\eta_0 = b^n / a^n \quad . \qquad (7)$$

The s-wave elastic scattering trial wavefunction for the system may be written in abbreviated form as:

$$\Psi_t^n = S + K_{11}\, C + \phi_n \qquad (8)$$

where S is the regular part ;

$$S = \hat{S}\,.\,\Phi_T\,(r) = \frac{1}{\sqrt{4\pi}} \sin c\beta \,.\, \Phi_T\,(r) \quad , \qquad (9)$$

($\sin c\,\beta = \dfrac{\sin\beta}{\beta}$ , $\beta = (k\,x)$ *where k* is the momentum of the incident positron). The function C consists of a cut-off function and the irregular part, i.e.

$$C = (1 - e^{-\alpha x})\hat{C}\Phi_T\,(r) \qquad (10)$$

$$= \frac{1}{\sqrt{4\pi}}\,(1 - e^{-\alpha x})(\cos c\beta)\,\Phi_T\,(r) \quad , \qquad (11)$$

where $\cos c\beta = \dfrac{\cos\beta}{\beta}$ *and* $\alpha$ is an adjustable (free) parameter which is selected from the values that give a plateau of $K_{11}$ (see ref. [11], P.73). $\Phi_T(r)$ is the target ground state wavefunction (see Appendix). The square integrable $\phi_n(x, r)$ possesses the form

$$\phi_n(x,r) = \Phi_T(r).\sum_{i=1}^{n} d_i\, \chi_i(x) = \sum_{i=1}^{n} d_i \phi_i \qquad (12)$$

where $\chi_i = x^i\, e^{-\alpha x}$ and $\phi_i = \chi_i\, \Phi_T$ . (13)

The next step in the variational treatment is to select a proper test-wave function $\phi_S$ and define the functional

$$\left\langle \phi_S \,\middle|\, H - E \,\middle|\, \Psi_t^n \right\rangle = V \qquad (14)$$

The linear variational parameters $K_{11}$ and $d_i$ are chosen according to the following variational principle:

$$\delta\, |V|^2 = 0 \qquad (15)$$

Thus, they are chosen following a least-squares variational principle in which the squared modulus of the projection of the vector $(H-E)\Psi_t^n$ in $\phi_S$ is minimal. The test wavefunction $\phi_S$ is constructed [11] by:

$$\phi_S = \{ S\ , C\ , \phi_j\ ;\ j = 1, 2, .....n \}. \qquad (16)$$

In this case we have the system of projections

$$\left(S\ |S\right)+K_{11}\left(S\ |C\right)+\sum_{i=1}^{n}d_i\ \left(S\ |\phi_i\right)=V_1$$

$$\left(C\ |S\right)+K_{11}\left(C\ |C\right)+\sum_{i=1}^{n}d_i\ \left(C\ |\phi_i\right)=V_2 \qquad (17)$$

$$\left(\phi_j\ |S\right)+K_{11}\left(\phi_j\ \ |C\right)+\sum_{i=1}^{n}d_i\ \left(\phi_j\ |\phi_i\right)=V_{j+2}\ ;\ j=1,2,....n\,.$$

The LSVM implies:

$$\delta\ \sum_{j=1}^{n+2}\left|V\ \right|_j^2 = 0\ . \qquad (18)$$

This means that the sum of squared moduli of the projections of $\left(H-E\right)\Psi_t^n$ on the test function space $\phi_s$ is minimum.

The minimization of $\sum_{j=1}^{n+2}\left|V\right|_j^2$ guarantees that the vector $\left(H-E\right)\Psi_t^n$ has a minimum length. The variational parameters are obtained by applying this variational principle (18). The total Hamiltonian (in Rydberg units) of positron-target atom system has the form:

$$H = H_T - \nabla_x^2 + V_{int}(r,x) \qquad (19)$$

where $H_T$ is the Hamiltonian of the target atom, $\nabla_x^2$ is the kinetic energy operator for the incident positron, $V_{int}(r,x)$ stands for the interaction potential between the positron and the target and r is used to represent the assemblage coordinate for Z atomic electrons.

The total energy E of the system may be written, in Rydberg, as

$$E = E_T + k^2 \ , \qquad (20)$$

where $E_T$ *and* $k^2$ are the energy of the target and the kinetic energy of the incident positrons, respectively. $V_{int}(r, x)$ is the interaction potential between the incident positron and the target and is given by

$$V_{int}(r, x) = \frac{2Z}{x} - \sum_{i}^{N} \frac{2}{|x - r_i|} \qquad (21)$$

Thus, the final form of the trial expansion space $\Psi_t^n$ can be expressed (see Appendix) in terms of wavefunction determinants as

$$\Psi_t^n = \frac{1}{\Delta : n} \left\{ \left| \begin{matrix} S & \phi_1 \ldots \phi_n . \\ M_:^S & \Delta : n \end{matrix} \right| + K_{11} \left| \begin{matrix} C & \phi_1 \ldots \phi_n . \\ M_:^C & \Delta : n \end{matrix} \right| \right\} \qquad \textbf{(2}2)$$

$Z_{eff}$ can be determined experimentally to a high degree of accuracy and thus, the calculation of this parameter is a criterion for the goodness of the employed wavefunction $\Psi_t$ which represents the system of a low energy positron moving in the field of the atom and approximated by eq.(8). Using equations defined for $\Psi_t$, $Z_{eff}$ can be written as

$$Z_{eff}(k) = \int \Psi_t^2 \ \delta \ (r - x) dr. \qquad (23)$$

Therefore, we have

$$Z_{eff}(k) = \frac{1}{4\pi} \int dr \ [ \ \Psi_P \ (r ; k) \ ]^2 \sum_{j} \left( R_{n_j \ell_j} \right)^2 \qquad (24)$$

or

$$Z_{eff}(k) = \frac{1}{4\pi} \int dr \left[ \sin ckr + K_{11} \cos ckr \left(1 - e^{-\alpha r}\right) + \sum_{i=1}^{n} d_i e^{-\alpha r} r^i \right]^2 \sum_j \left(R_{n_j \ell_j}\right)^2 \quad (25)$$

where $K_{n_j \ell_j}$ denotes the electron radial wavefunctions and the summation is over the electron states in the atomic level defined by quantum numbers $n_j$ *and* $\ell_j$ ( see Appendix ).

The s-wave elastic scattering cross section (in $\pi a_0^2$ units) is related to the phase shift by the following relation

$$\sigma_{el} = \frac{4}{k^2} \sin^2(\eta_0). \quad (26)$$

# 3-Results and discussion

The computation of the annihilation rates was started by calculating the orbital wavefunctions and energies of the target atoms using Cowan computer code (program RCN32). These wavefunctions were used for calculating the positron-atom potentials (eq. A5b). After the construction of the matrix elements $(S|S)$, $(S|C)$, $(S|\phi_i)$, $(C|S)$, $(C|C)$, $(C|\phi_i)$, $(\phi_j|S)$, $(\phi_j|C)$, and $(\phi_j|\phi_i)$, we employed the LSVM program (at certain starting values of the free non linear parameter $\alpha$ and n (the number of the square integrable functions included in the trial wavefunction) in order to test the validity of the whole program. Later on, we changed $\alpha$ and increased n until we reached convergence and stability in the results of $K_{11}$. This was achieved at $\alpha$ = 0.3 and n = 7.

Our computational process of the annihilation rates started with calculations of this quantity argon atoms. These results were compared with already existing experimental data, see Fig.(1). This figure shows that our theoretical calculations have a good agreement with the experimental data and show similar behavior as the results of Mitroy and Ivanov [12] developed using a two-parameter semiempirical theory of positron scattering and annihilation. Obviously, our results lie higher than the value $Z_{eff}$ = 13.6 obtained by Dzuba et al [13]. The improvements demonstrated by our calculations are attributed to the accurate forms of the bound state wavefunctions and the positron wavefunctions obtained, respectively, via ab initio and least-squares variational techniques.

Since the ionization energies of the argon, iron and copper atoms are approximately 15.76 eV, 7.87 eV and 7.726 eV, respectively, the energy of the incident positron $k^2$ must lie below the so called Ore gap, i.e. it must be less than the difference between the values of the ionization energies and the binding energy of the Ps (- 6.8 eV). This means that $k^2$ should be less than the Ps formation thresholds 8.96 eV, 1.07 eV and 0.926 eV, respectively, where the only possible channels are the elastic scattering and the direct annihilation. Therefore, we have calculated the effective charge ( $Z_{eff}$ ) at each subshell in the collision of positrons with Ar, Fe and Cu atoms through the energy ranges below 8.96 eV, 1.07 eV and 0.926 eV, respectively.

The contribution of each subshell to the total effective charge for the collision of positrons with argon, iron and copper atoms are shown in figures (2), (3) and (4), respectively. The annihilation parameters for iron and copper atoms are plotted in figures (5) and (6). Fig.6 contains also a comparison with the results of Mitroy and Ivanov [12] . The s-wave elastic scattering cross-sections $\sigma_{el}$ $(in\, units\, of\, \pi a_0^2)$ of positrons by iron and copper atoms are drawn in figures (7) and (8). These figures demonstrate the monotonic decrease of $\sigma_{el}$ as the energy of the incident positron increases.

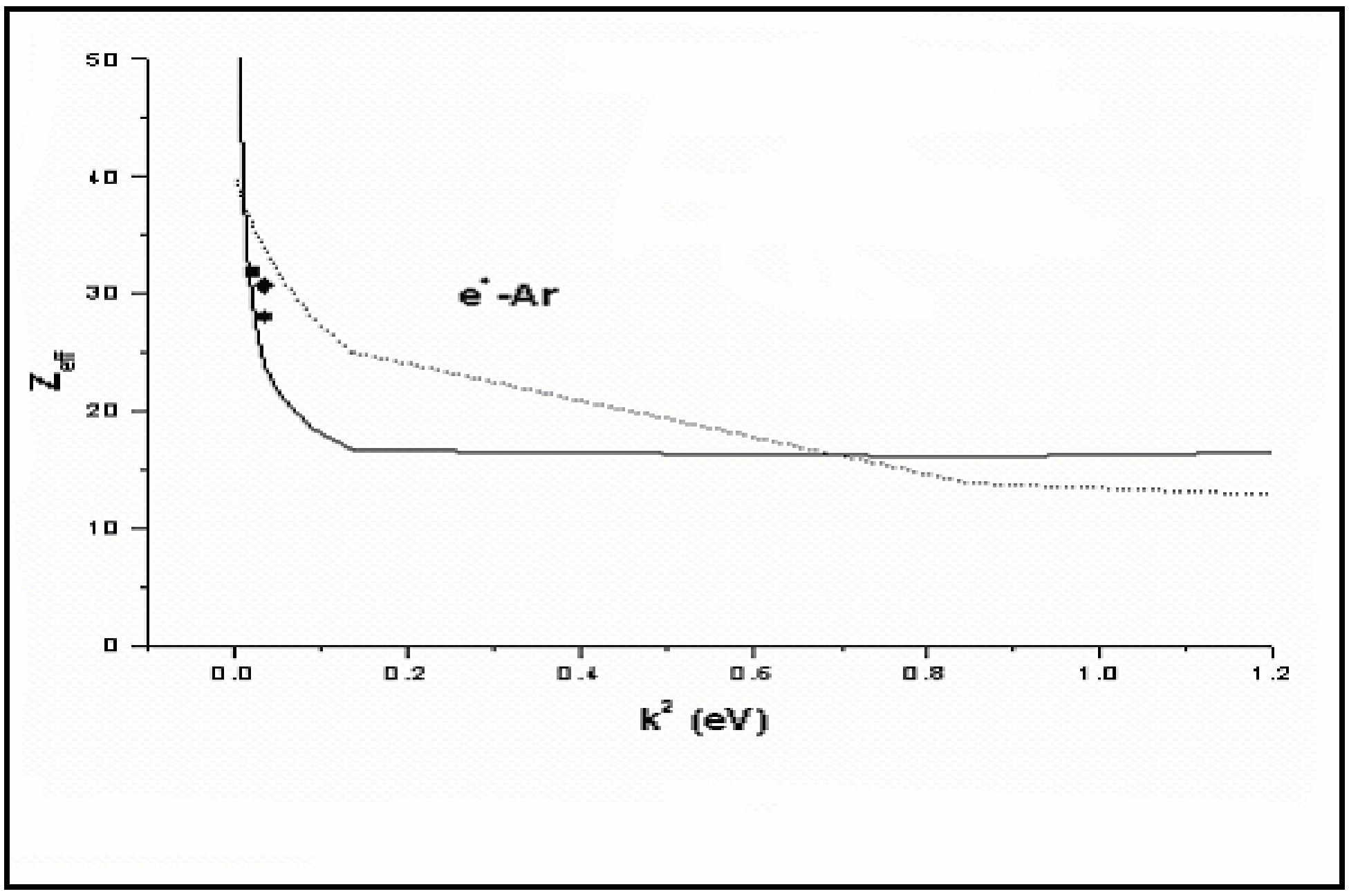

**Fig.1: Comparison between present ( — ) annihilation parameter ($Z_{eff}$) of $e^+$- Ar scattering and the results of Mitroy and Ivanov ( - - - ) [12] Canter and Roellig ( ■ ) [14], and Paul ( ♦ ) [15].**

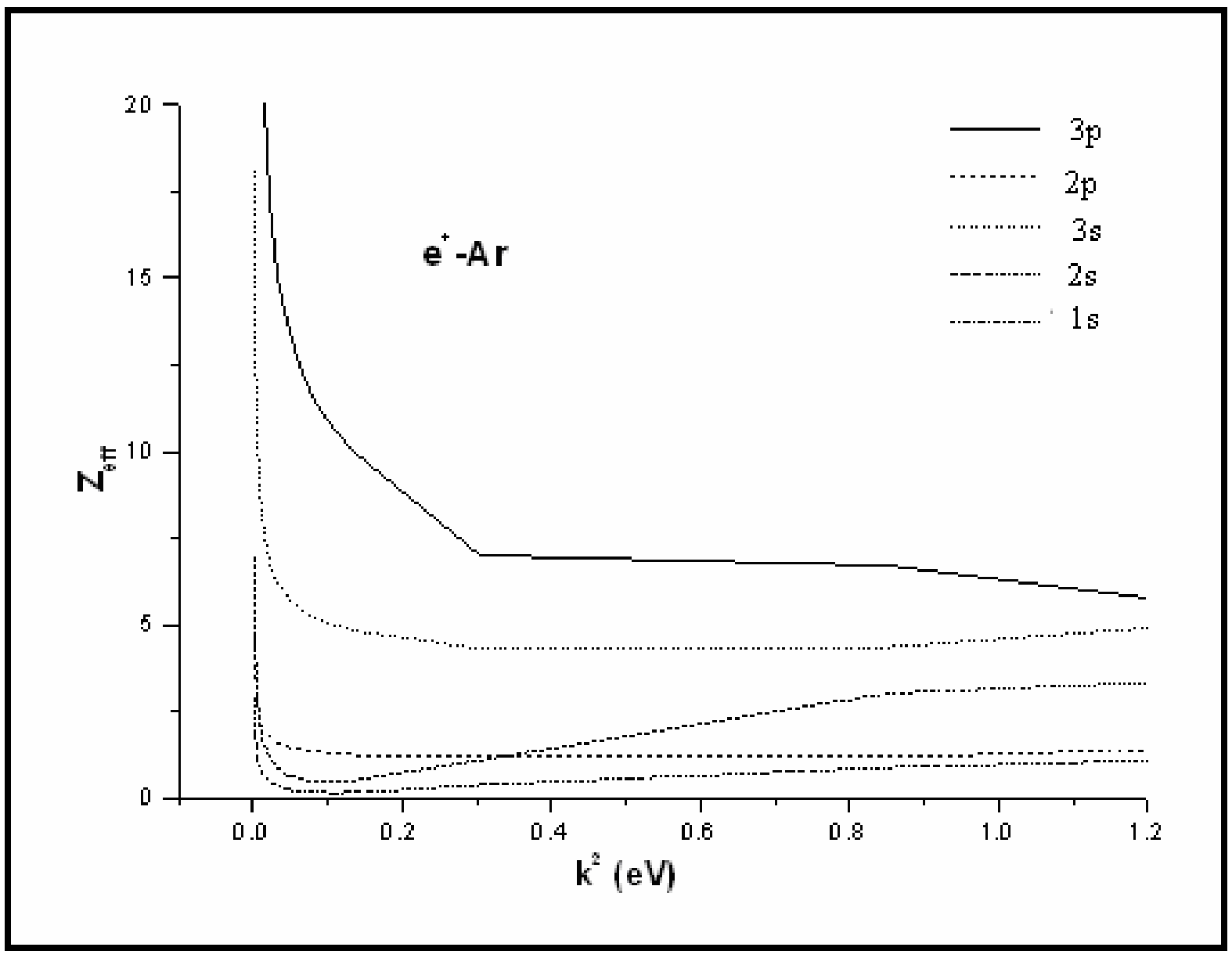

**Fig.2: The annihilation factor ($Z_{eff}$) as a function of the incident positron energy ($k^2$) for different subshells of positron-argon scattering.**

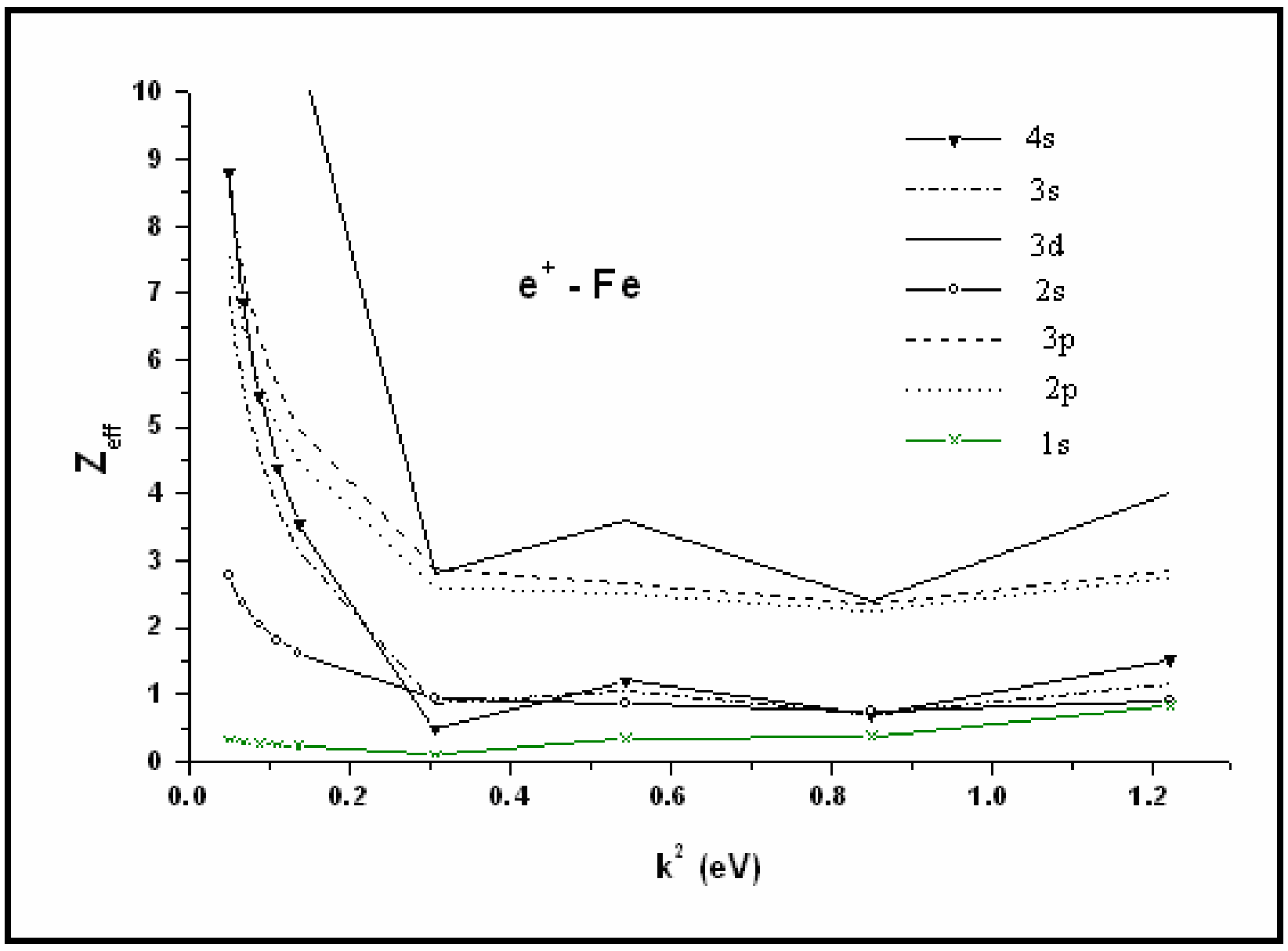

**Fig.3. Energy dependence of the annihilation parameter ($Z_{eff}$) for different subshells of positron-iron scattering.**

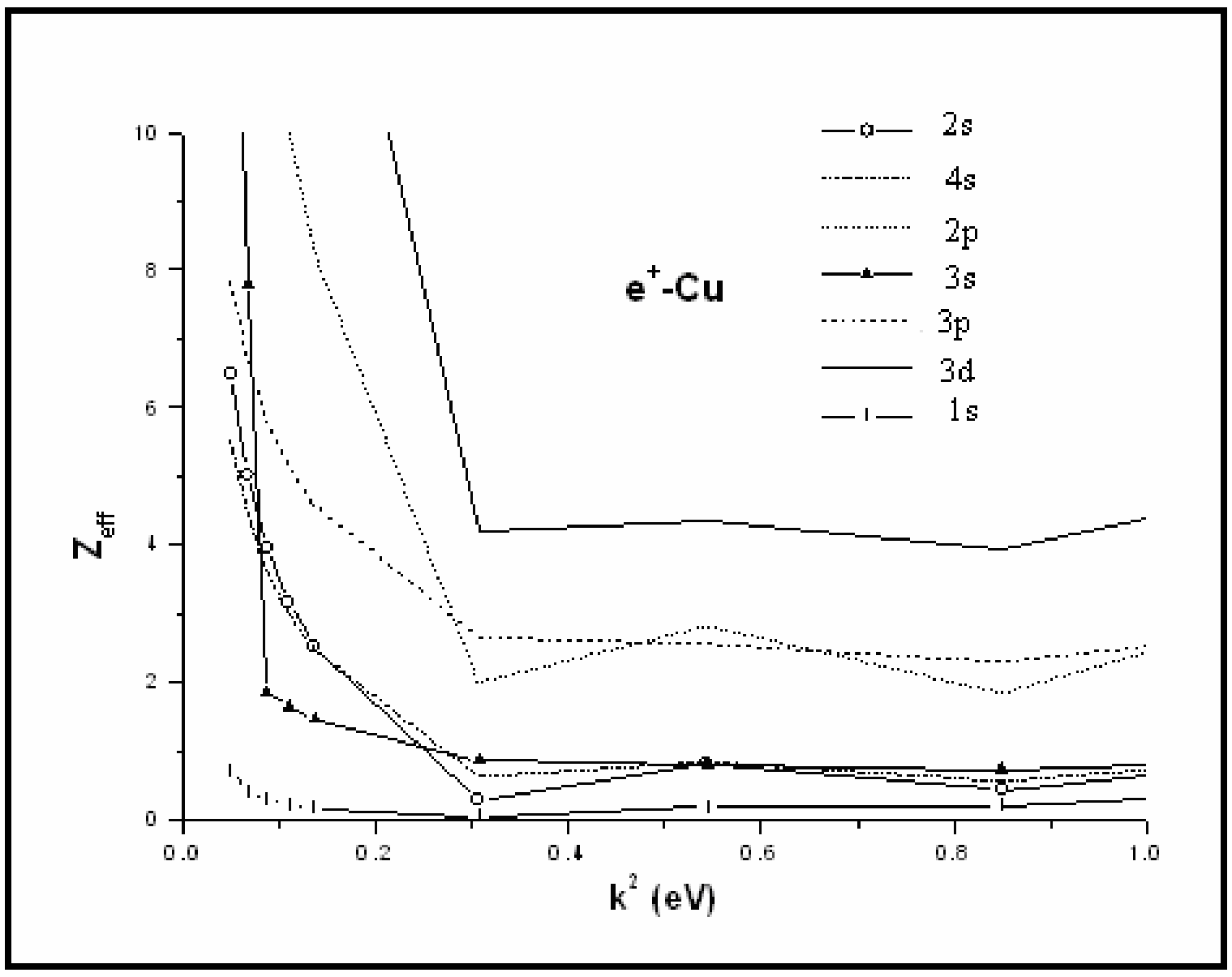

**Fig.4.The energy dependence of the annihilation parameter ($Z_{eff}$) for different subshells of positron-copper scattering.**

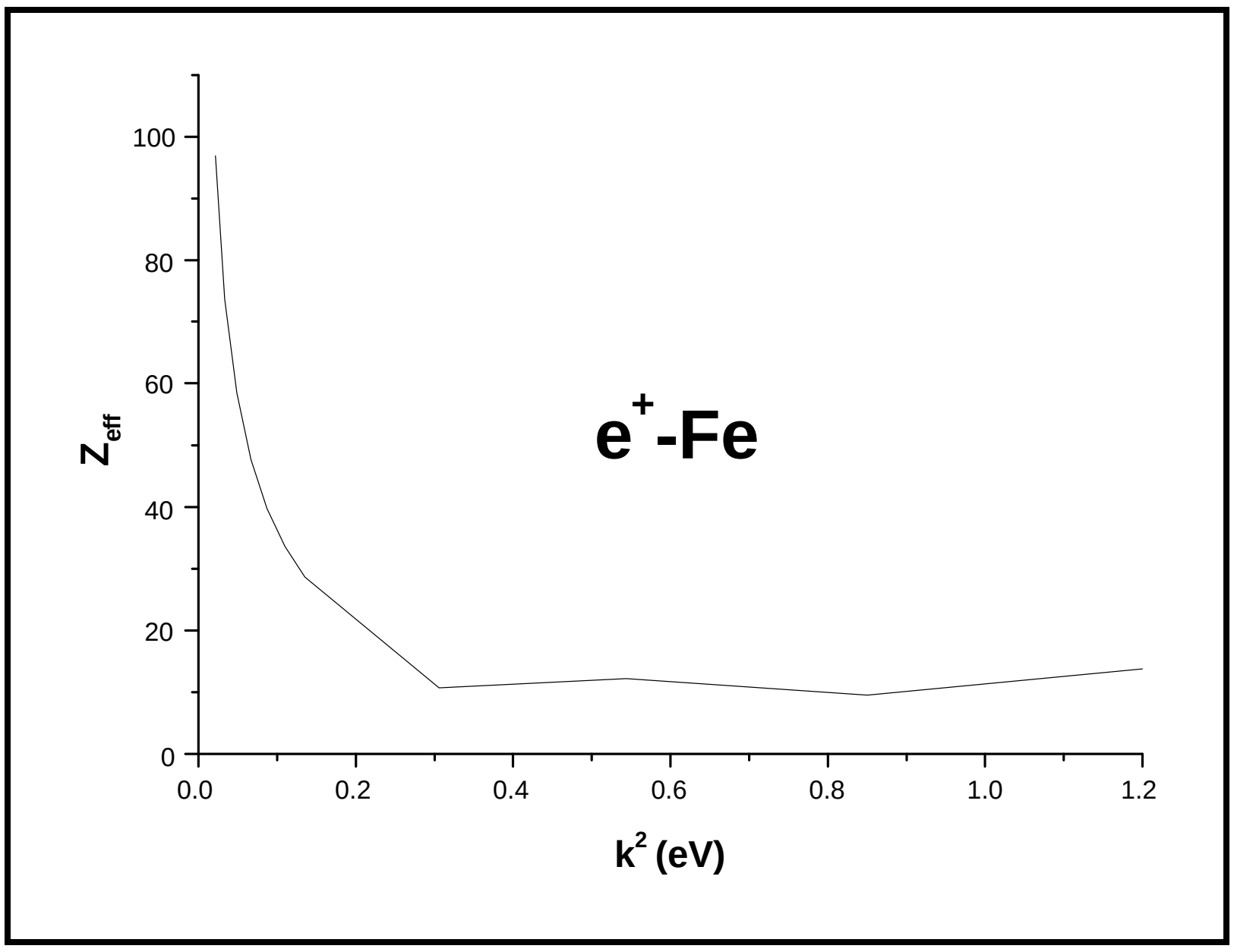

**Fig. 5: Total effective charge ($Z_{eff}$) as a function of the incident positron energy ($k^2$) for positron-iron scattering.**

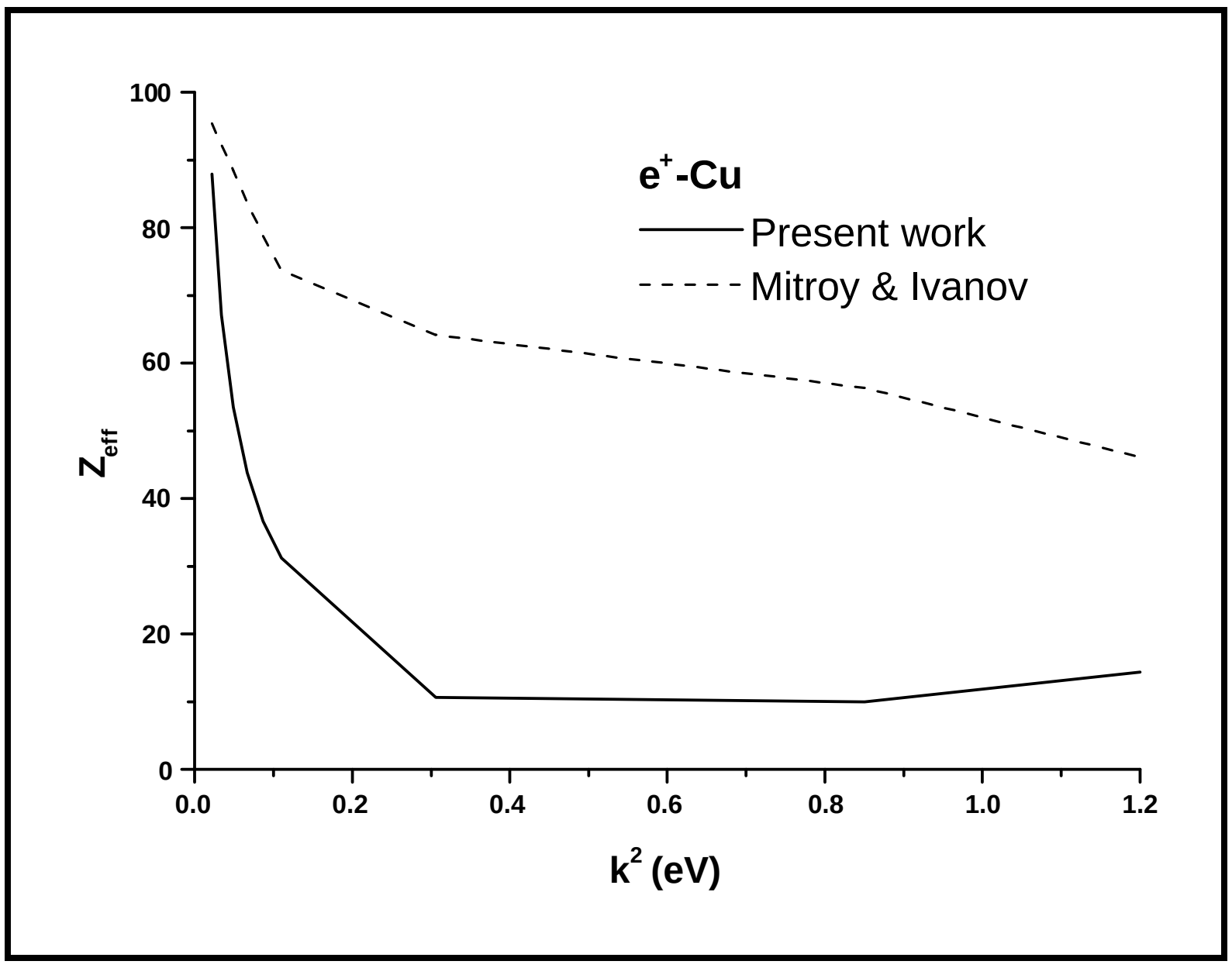

**Fig.6: The total effective charge ($Z_{eff}$) as a function of the incident positron energy ($k^2$) for positron-copper scattering.**

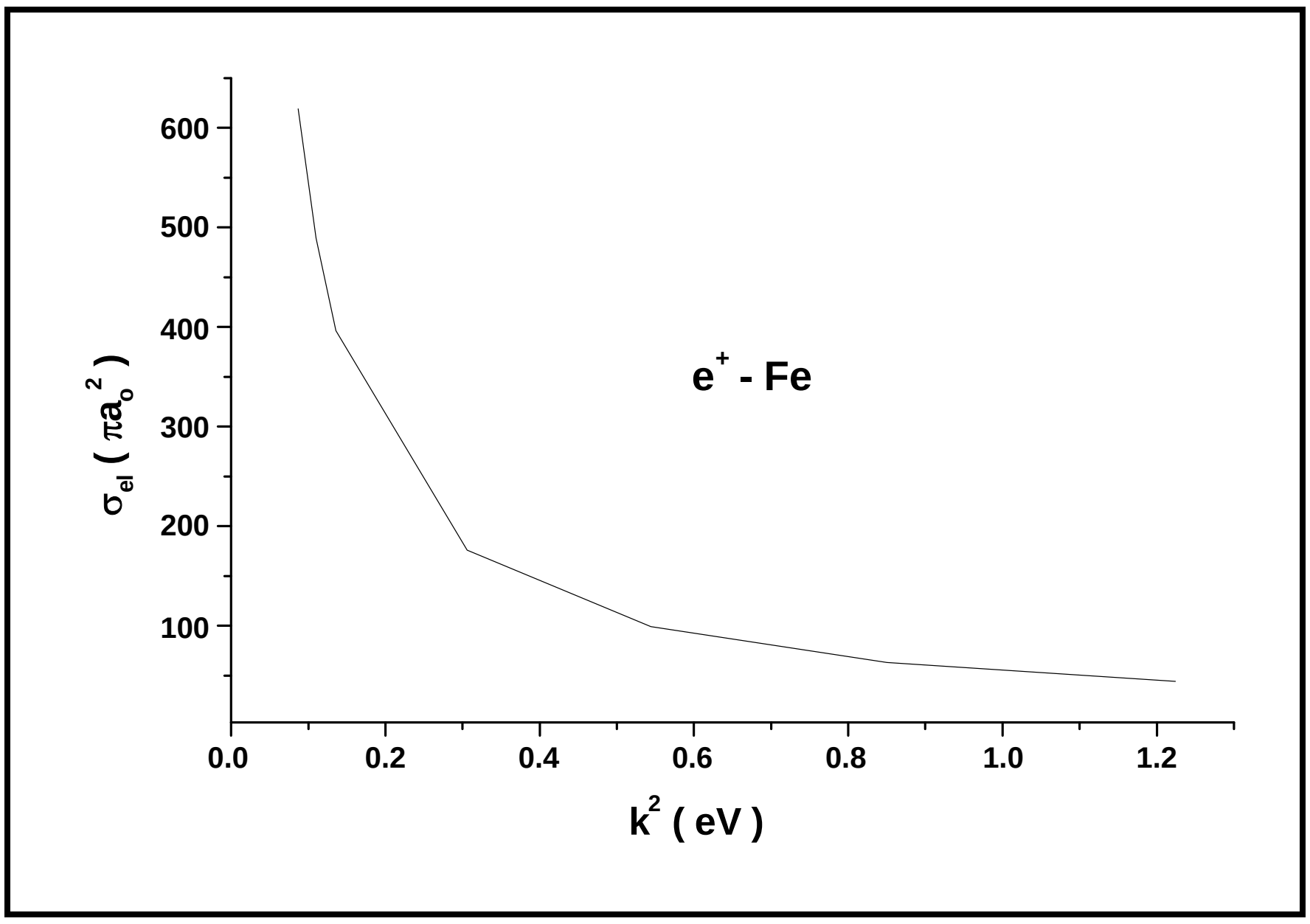

**Fig.7: The s-wave elastic cross-sections for positron-iron scattering.**

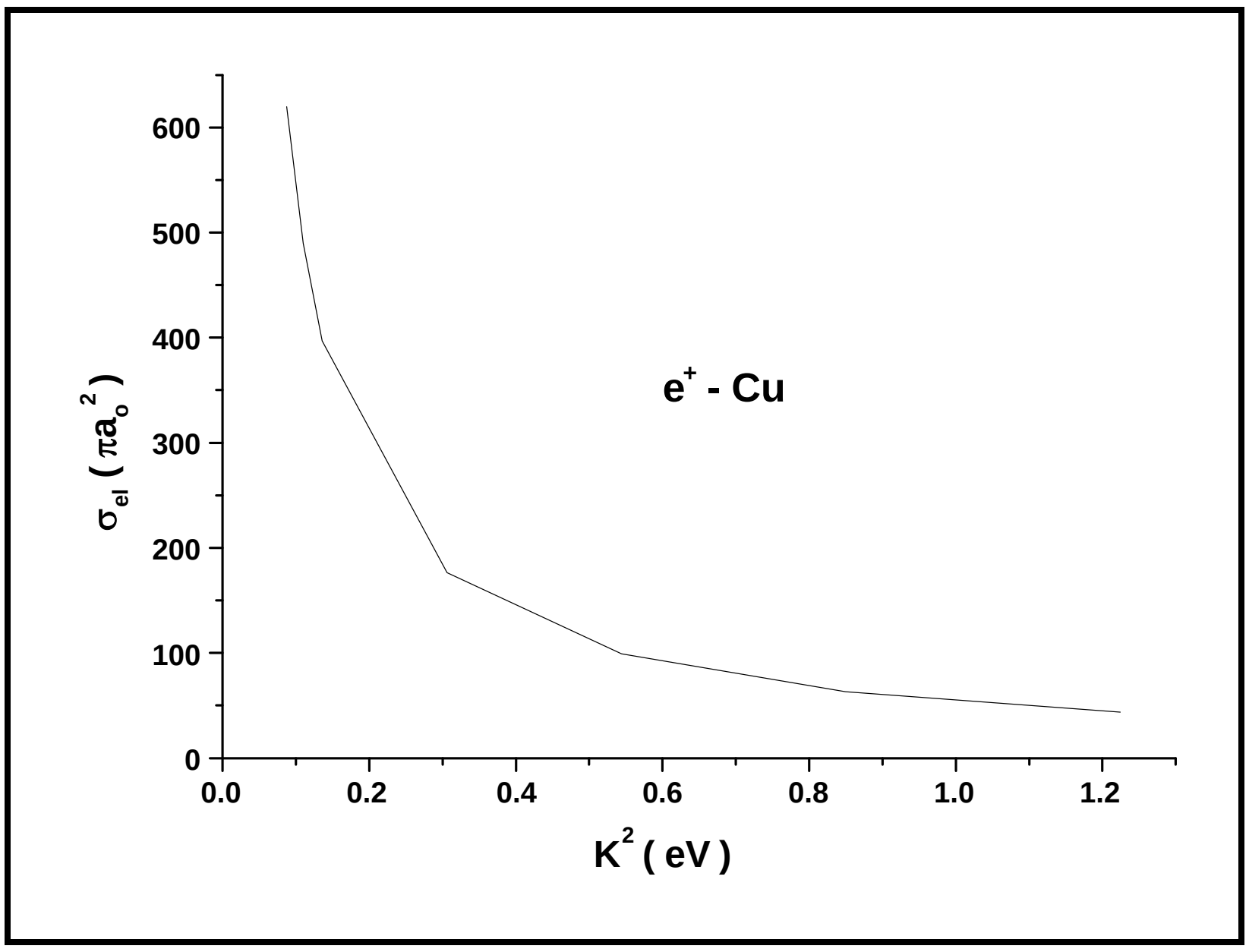

**Fig.8: The s-wave elastic cross-sections for positron-copper scattering.**

## Appendix

This appendix contains a brief discussion of the potential of the positron in the target field and the matrix elements needed to specify the matrices Q and q required for calculating the variational parameters using the least-squares variational method (LSVM) program.

The target ground state wavefunction can be expressed as a Slater determinant of mutually orthonormal one-electron wavefunctions $u_i$ in the form:

$$\Phi_T\,(r) \;= \frac{1}{\sqrt{z!}}\,\det\left[\,u_1(r_1)\;u_2(r_2)\;u_3(r_3)\;.....u_z(r_z)\;\right] \tag{A1}$$

In equation (A1) z denotes the total number of electrons. According to the central field model (Heyland et al [16]), $u_i\,(r_i)$ can be expressed as

$$u_i\,(r_i) \;=\; \frac{1}{r_i}\,R_{n_i\,\ell_i}\,Y_{\ell_i m_i}\,(\hat{r}_i)\;\zeta\,(\sigma\,)\,, \tag{A2}$$

where $R_{n_i\,\ell_i}$ is the radial wavefunction, $Y_{\ell_i m_i}\,(\hat{r}_i)$ are the usual spherical harmonics and $\zeta\,(\sigma\,)$ stands for the spin vector of the orbit i such that $n_i\,,\;\ell_i$ *and* $m_i$ are the corresponding principal, orbital and magnetic quantum numbers, respectively. The energy of the target is given by

$$E_T \;= \int \Phi_T(r)\;H_T\;\Phi_T(r)\;dr \tag{A3}$$

In order to calculate the matrix elements, the positron potential has to be determined. The potential $U(x)$ of the positron in the target field is defined as

$$U(x) = \left\langle \Phi_T(r) \,\middle|\, V_{int}(r,x) \,\middle|\, \Phi_T(r) \right\rangle\;. \tag{A4}$$

In the original work of Madison and Shelton [17] $U(x)$ takes the form:

$$U(x) = \sum_i w_i \left\{ \frac{2Z}{x} \int_0^{\infty} dr_i \, R^2_{n_i \ell_i} - 2 \int_0^{\infty} dr_i \frac{R^2_{n_i \ell_i}}{\max(x, r_i)} \right\} \tag{A5a}$$

It can be written (Cowan 1981) as:

$$U(x) = \sum_i w_i \left\{ \frac{2Z}{x} \int_0^{\infty} dr_i \, R^2_{n_i \ell_i} - \frac{2}{x} \int_0^{x} dr_i R^2_{n_i \ell_i} - 2 \int_x^{\infty} dr_i \frac{R^2_{n_i \ell_i}}{r_i} \right\} \tag{A5b}$$

where $w_i$ denotes the occupation number of electrons in $n_i \, \ell_i$ atomic orbit and $n_i, \ell_i$ are the principal and orbital quantum numbers of an orbit $i$, respectively. The orbital radial wavefunction $R_{n_i \ell_i}$ is the solution of the equation:

$$\left[ -\frac{d^2}{dr_i^2} + \frac{\ell_i(\ell_i + 1)}{r_i^2} + V_i(r_i) \right] R_{n_i \ell_i} = \varepsilon_i \, R_{n_i \ell_i} \tag{A6}$$

where $V_i(r_i)$ is the assumed potential energy function for the field in which the atomic electron $i$ moves. These functions are generated from Cowan program, which is based on the description of Herman and Skillman [18] with Hartree plus statistical exchange approximated potential.

The system of (n + 2) equations, i.e. eqs.(17),can be reduced in matrix representation to the form:

$$Q\,d - q = V \tag{A7a}$$

where the matrices Q , d , q and V are defined below. In other words, the least-squares principle is equivalent to the minimization of the norm of the vector $Q\,d - q$ , which leaves us with

$$Q^+ Q\ d = Q^+ q$$
$$d = \left(Q^+ Q\right)^{-1} Q^+ q \qquad \text{(A7b)}$$

$$Q = \begin{pmatrix} (S|C) & M^{ST} \\ (C|C) & M^{CT} \\ M^C & \Delta_n \end{pmatrix},\ d = \begin{pmatrix} R_{11} \\ d_1 \\ \vdots \\ d_n \end{pmatrix},\ q = -\begin{pmatrix} (S|S) \\ (C|S) \\ M^S \end{pmatrix} \text{and } V = \begin{pmatrix} V_1 \\ V_2 \\ \vdots \\ V_{n+2} \end{pmatrix} \qquad \text{(A8)}$$

where $M^{ST}$ *and* $M^{CT}$ are the transpose of the column vector $M^S$ *and* $M^C$ ,respectively, which are defined as

$$M^S = \begin{pmatrix} (\phi_1|S) \\ (\phi_2|S) \\ \vdots \\ (\phi_n|S) \end{pmatrix},\ M^C = \begin{pmatrix} (\phi_1|C) \\ (\phi_2|C) \\ \vdots \\ (\phi_n C) \end{pmatrix} \text{ and } \Delta_n = \begin{pmatrix} (\phi_1|\phi_1) & \cdots & (\phi_1|\phi_n) \\ \cdots & \cdots & \cdots \\ (\phi_n|\phi_1) & \cdots & (\phi_n|\phi_n) \end{pmatrix} \qquad \text{(A9)}$$

The closed form of the matrix elements required for the employment of the LSVM, namely $(S\,|S)$, $(S\,|C)$, $(S\,|\phi_i)$ ,$(C\,|S)$, $(C\,|C)$, $(C\,|\phi_i)$, $(\phi_j\,|S)$, $(\phi_j\,|C)$, and $(\phi_j\,|\phi_i)$ , are needed. These matrix elements have the general form:

$$(g\,|\,f\,) = \langle g\,|\,E-H\,|\,f\,\rangle = \int_0^\pi \sin\theta\,d\theta \int_0^{2\pi} d\varphi \int_0^\infty x^2 g\hat{H}f\ dx \qquad \text{(A10)}$$

where x is the position vector of the positron, $\theta$ is the angle between x and the Z-axis and $\varphi$ is the azimuthal angle. The operator $\hat{H}$ possesses the form

$$\hat{H} = (E - H),$$ (A11)

which can be written in several different ways depending on the particular form of wavefunction on which it operates.

The effects of $\hat{H}$ *on* $S, C$ *and* $\phi_i$ are given by:

$$\hat{H}\ S = \Phi_T(r)\left[-V_{int}(r,x)\right]\sin c\beta \Big/ \sqrt{4\pi}$$ (A12a)

$$\hat{H}\,C = \Phi_T(r)\left[-V_{int}(r,x)\frac{1`}{\sqrt{4\pi}}\left(1-e^{-\alpha x}\right)\cos c\beta - 2\alpha\, k\, e^{-\alpha x}\sin c\beta - \alpha^2 e^{-\alpha x}\cos c\beta\right]$$

(A12b)

$$\hat{H}\,\phi_i = \Phi_T(r)\left[-V_{int}(r,x)\ x^i e^{-\alpha x} + \left(k^2+\alpha^2\right)x^i e^{-\alpha x} - 2\alpha\left(i+1\right)x^{i-1}e^{-\alpha x} + i\left(i+1\right)x^{i-2}e^{-\alpha x}\right]$$

(A12c)

(Remember that $\sin c\beta = \dfrac{\sin\beta}{\beta}$, $\beta = (k\,x)$ and $\cos c\beta = \dfrac{\cos\beta}{\beta}$ ).

Therefore, the matrix elements appearing in the matrices Q and q have the following final forms:

$$\left(S\,|S\right) = -\frac{1}{k^2}\int_0^\infty U(x)\ \sin^2\beta\, dx$$ (A13)

$$\left(S\,|C\right) = -\frac{1}{k^2}\int_0^\infty\left[U(x)\sin\beta\cos\beta\,(1-e^{-\alpha x})\right]dx - \frac{1}{k} + \frac{\alpha\ \cos\delta}{A^{1/2}} - L\frac{\sin\delta}{A^{1/2}}$$ (A14)

$$\left(C\,|S\right) = -\frac{1}{k^2}\int_0^\infty\left[U(x)\sin\beta\cos\beta\,(1-e^{-\alpha x})\right]dx$$ (A15)

$$\left(C\,|C\right) = -\frac{1}{k^2}\int_0^\infty \left[U(x)\cos^2\beta\ (1-e^{-\alpha x})\right]dx - M\ \frac{\sin\delta}{A^{1/2}}$$

$$+M\ \frac{\sin\varepsilon}{B^{1/2}} + L\ \frac{\cos\delta}{A^{1/2}} - L\ \frac{\cos\varepsilon}{B^{1/2}}. \qquad \text{(A16)}$$

$$\left(S\,|\phi_i\right) = -\frac{1}{k}\int_0^\infty \left[U(x)\sin\beta\ x^{(i+1)}\ e^{-\alpha x}\right]dx + \frac{(i+1)!}{k\,D^{i/2}}\left[\ \sin((i+2)\varepsilon) - \frac{2\alpha\ \sin((i+1)\varepsilon)}{F^{1/2}} + \sin(i\varepsilon)\right]$$

.

(A17)

The elements of $M^{CT}$ are given by

$$\left(C|\phi_i\right) = -\frac{1}{k}\int_0^\infty \left[U(x)\cos\beta\ x^{(i+1)}(1-e^{-\alpha x})\right]dx + \frac{(i+1)!}{k\,D^{i/2}}\left[\ \cos((i+2)\varepsilon) - \frac{2\alpha\ \cos((i+1)\varepsilon)}{D^{1/2}} + \cos(i\varepsilon)\right]$$

$$+\frac{(i+1)!}{k\,G^{i/2}}\left[\frac{D}{G}\ \cos((i+2)\zeta) + \frac{2\alpha\ \cos((i+1)\zeta)}{D^{1/2}} - \cos(i\zeta)\right]$$

(A18)

$$\left(\phi_j\,|S\right) = -\frac{1}{k}\int_0^\infty U(x)\sin\ \beta\left[e^{-\alpha x}\,x^{(j+1)}\right]dx \qquad \text{(A19)}$$

$$\left(\phi_j\,|C\right) = -\frac{1}{k}\int_0^\infty U(x)\ x^{(j+1)}e^{-\alpha x}(1-e^{-\alpha x})\cos\ \beta\ dx - \frac{(j+1)!}{G^{\frac{j}{2}+1}}\left[2\alpha\sin((j+2)\varepsilon) - \frac{\alpha^2}{k}\cos((j+2)\zeta)\right]$$

(A20)

$$\left(\phi_j\,|\phi_i\right) = -\int_0^\infty U(x)e^{-\alpha x}\,x^{(j+i+2)}\ dx + D\frac{(j+i+2)!}{2\alpha^{j+i+3}} - 2\alpha(i+1)\frac{(j+i+1)!}{2\alpha^{j+i+2}} + i(i+1)\frac{(j+i)!}{2\alpha^{j+i+1}}$$

(A21)

where

$$\delta = \arctan\left(\frac{2k}{\alpha}\right),\ \varepsilon = \arctan\left(\frac{k}{\alpha}\right),\ \zeta = \arctan\left(\frac{k}{2\alpha}\right), A = \left(\alpha^2 + 4k^2\right), B = \left(4\alpha^2 + 4k^2\right),$$

$$D = \left(\alpha^2 + k^2\right), F = \left(-\alpha^2 + k^2\right), G = \left(4\alpha^2 + k^2\right), L = \left(\alpha^2/2k^2\right)\ and\ M = \left(\alpha/k\right).$$

The final form of the trial expansion space $\Psi_t^n$ can be expressed in terms of vector determinants as

$$\Psi_t^n = \frac{1}{\Delta:n}\left\{ \begin{vmatrix} S & \phi .....\phi_n . \\ M_:^S & \Delta:n \end{vmatrix} + K_{11} \begin{vmatrix} C & \phi_1 ....\phi_n . \\ M_:^C & \Delta:n \end{vmatrix} \right\}$$

(A22)

where

$$M_:^S = \begin{pmatrix} (\phi_1 : S) \\ (\phi_2 : S) \\ \vdots \\ (\phi_n : S) \end{pmatrix}, \; M_:^C = \begin{pmatrix} (\phi_1 : C) \\ (\phi_2 : C) \\ \vdots \\ (\phi_n : C) \end{pmatrix} \text{ and } \Delta:n = \begin{pmatrix} (\phi_1 : \phi_1) & (\phi_1 : \phi_2) & \cdots & (\phi_1 : \phi_n) \\ (\phi_2 : \phi_1) & (\phi_2 : \phi_2) & \cdots & (\phi_2 : \phi_n) \\ \cdots & \cdots & \cdots & \cdots \\ (\phi_n : \phi_1) & (\phi_n : \phi_2) & \cdots & (\phi_n : \phi_n) \end{pmatrix}$$

(A23)

The elements of the Hermitean matrix $Q^+ Q$ can be abbreviated by $( f:g )$. The matrix element $( f:g )$ is defined by $( f:g ) = \sum_{k=1}^{n} ( f|\phi_k )(\phi_k |g)$ . The positron wavefunction $\Psi_P^n$ can be also expressed in terms of vector determinants as

$$\Psi_P^n = \frac{1}{\Delta:n}\left\{ \begin{vmatrix} \hat{S} & \chi_1 .....\chi_n \\ M_:^S & \Delta:n \end{vmatrix} + K_{11} \begin{vmatrix} \hat{C} & \chi_1 .....\chi_n \\ M_:^C & \Delta:n \end{vmatrix} \right\}$$

(A24)

In the final analysis, $K_{11}$ is given in the following form

$$K_{11} = -\Delta_n^{(C:S)} / \Delta_n^{(C:C)} \qquad \text{(A25)}$$

where the determinants $\Delta_n^{(f:g)}$ are defined as

$$\Delta_n^{(f:g)} = \begin{vmatrix} ( f:g ) & M^{f\,T} \\ M^g & \Delta_n \end{vmatrix} \qquad \text{(A26)}$$

where $M^{f\,T}$ is the transpose of $M^f$ and $\begin{matrix} f \\ g \end{matrix} \equiv S \; or C$.